\documentclass[5p]{elsarticle}

\usepackage{lineno,hyperref}
\usepackage[T1]{fontenc}
\usepackage{booktabs}
\usepackage{mathptmx}
\usepackage{graphics}
\usepackage{multirow}
\usepackage{subfigure}
\usepackage{graphicx}
\usepackage{amsmath}
\usepackage{makecell}
\usepackage{tablefootnote}
\usepackage{threeparttable}

\modulolinenumbers[5]

\journal{Journal of \LaTeX\ Templates}
\begin{document}
\title{\textbf{Crystal structures and electronic and magnetic properties of Janus bilayer Cl$_3$Cr$_2$I$_3$ }}%

\author[1]{Suqi Liu}%
\address[1]{School of Physics, Communication and Electronics, Jiangxi Normal University, Nanchang 330022, China}

\author[3]{Feng Sun}%
\address[3]{School of Education, Nanchang Institute of Science and Technology, Nanchang 330108, China}
\author[1]{Aijun Hong\corref{cor1}}%
\ead{6312886haj@163.com or haj@jxnu.edu.cn}

\cortext[cor1]{Corresponding author} 

\begin{abstract}
Two-dimensional (2D) magnetic material CrI$_3$  has aroused extensive attention, because it could provide a new platform for investigating the relations between crystal structures and electronic and magnetic properties.
Here, we study crystal structures and electronic and magnetic properties of three configurations (Cl-Cl I-I and Cl-I) of Janus bilayer Cl$_{3}$Cr$_{2}$I$_{3}$ with two stacking orders (AB and AA$_{1/3}$) by using the first principles approach incorporating the spin-orbit coupling (SOC) effect and the dipole correction.
It is found that the spin polarization and the SOC effect can expand the lattice constant and the interlayer distance (ID) of the three configurations. Especially, the ID of the I-I configuration is 1  \AA ~larger than that of the Cl-Cl configuration.
The total energy calculation results show that the atomic configuration, the SOC effect and the stacking order play an important role in determining the magnetic ground states of the Janus layers.
Our results indicate the atomic configurations of the Janus bilayers are not conducive to the increase of critical temperature. Interestingly, the Cl-I configuration has a vertical dipole moment, and its AFM state has large spin splitting.
It is revealed that the non-periodic structure, the symmetry breaking of the average potential and the weak interlayer interaction lead to the vertical dipole moment and the abnormal AFM state that is not the most stable state.

\end{abstract}
\maketitle

\section{Introduction}
In condensed matter physics, since two-dimensional (2D) magnetic materials have
been successfully prepared in 2018, they have received extensive attention owing to
their exotic physical properties having possible applications in spintronics, magnetic
memories and topologically protected magnons \cite{r1,r2,r3,r4}. Especially, chromium triiodide
(CrI$_3$) exhibits layer dependent interesting  magnetic order \cite{r2}. Ferromagnetism
in bulk CrI$_3$ changes to antiferromagnetism in bilayer, and back to ferromagnetism in
the monolayer. Recent studies indicated that electric gating or magnetic fields can induce the transition
from antiferromagnet to ferromagnet in bilayer CrI$_3$. Thus, some engrossing
phenomena such as giant tunneling magnetoresistance \cite{r5,r6,r7}, gate tunable
magnetooptical Kerr effect \cite{r8,r9,r10} and large spin filering \cite{r35} were observed or predicted. More interestingly, by asymmetric chemical modification of conventional 2D materials, an emerging derivative of 2D materials called Jannus 2D materials was born. The most famous of them is non-magnetic Janus 2D transition metal chalcogenides (TMDs) that exhibit many novel physics. For instance, theoretical studies predicted the second harmonic generation response \cite{r11,r12} and the Rashba effect in Janus MoSSe monolayer \cite{r13,r14}. Certainly, the magnetic Janus 2D materials have
more interesting  physic nature. Theoretically, 2D Cr$_2$X$_3$S$_3$
Janus semiconductors have room temperature magnetisms by replacing one layer of X(Br or I) atoms with S atoms \cite{r36}.
Spin splitting exists in electric potential difference antiferromagnetism Mn$_2$ClF \cite{r34}. Moreover, it has been predicted that I$_3$Cr$_{2}$Br$_3$ monolayer has perpendicular magnetic anisotropy, however, I$_3$Cr$_2$Cl$_3$ monolayer has in-plane magnetic anisotropy \cite{r15}.

Besides, with the continuous progress in preparation technology of 2D materials, people have a
strong interest in the materials with double-layer thickness. Unconventional superconductivity
was experimentally observed in magic-angle graphene \cite{r16}, which implies the stacking
order has a large effect on physical and chemical properties of materials, and thus opens up a new research field. For example, the interlayer distances of bilayers XS$_2$ (X= Mo, Cr) change with large twist angles \cite{r17}. It is predicted that twist angles can induce extremely flat bands in bilayer $\alpha$-In$_2$Se$_3$ \cite{r18} and graphene \cite{r19}. Stacking order in bilayer CrX$_3$ (X = I, Br, Cl) can manipulate interlayer magnetism \cite{r20}.
However, there is little research on Janus Cr$_2$Cl$_3$I$_3$  bilayer. Therefore, crystal structure and electronic and magnetic properties of the Janus Cl$_3$Cr$_2$I$_3$ bilayer remain worth exploring, although the Cl$_3$Cr$_2$I$_3$ are explored in detail, in which the SOC effect is considered in electronic structure calculation rather than crystal
structure optimization and the AA$ _{ 1/3 } $ stacking order is not discussed \cite{r21}.

In this work we employ the first principles calculations with the SOC effect and the dipole correction to investigate the crystal structure and electronic and magnetic properties of Janus Cl$_3$Cr$_2$I$_3$ bilayer. We determine the stable structure with the lowest energy and the magnetic ground state of the three configurations (Cl-Cl, I-I and Cl-I) and discuss the relaxations between structures and electronic and magnetic
properties.

\section{Methodology}
The density functional theory (DFT) calculations were performed in Vienna ab initio simulation package (VASP) \cite{r22,r23} using the generalized gradient
approximation parametrized by the Perdew-Burke Ernzerhof (GGA-PBE) functional describing the exchange-correlation effects  \cite{r24,r25}. Sufficiently large energy cutoff of 600 eV was set for the plane-wave basis sets. For crystal structure optimization, we adopted the electronic convergence of 10$^{-6}$ eV, the ionic convergence of 0.01 eV/{\AA} and $\boldsymbol{k}$-mesh of 9$\times$9$\times$1 points. In order to avoid periodic interactions, a vacuum layer with a thickness of greater than 20 {\AA} is added to the models, and the length of the $c$-axis is fixed during the optimization process. The spin-orbit coupling effects and the dipole correction are included in all the calculations. The Bader charge analysis and the mean field approximation are used to calculate the dipole moment and the critical temperature, respectively.

\section{Results and discussion}
\subsection{Crystal structures}
\begin{figure}[h]
\centering
\includegraphics[width=0.8\columnwidth]{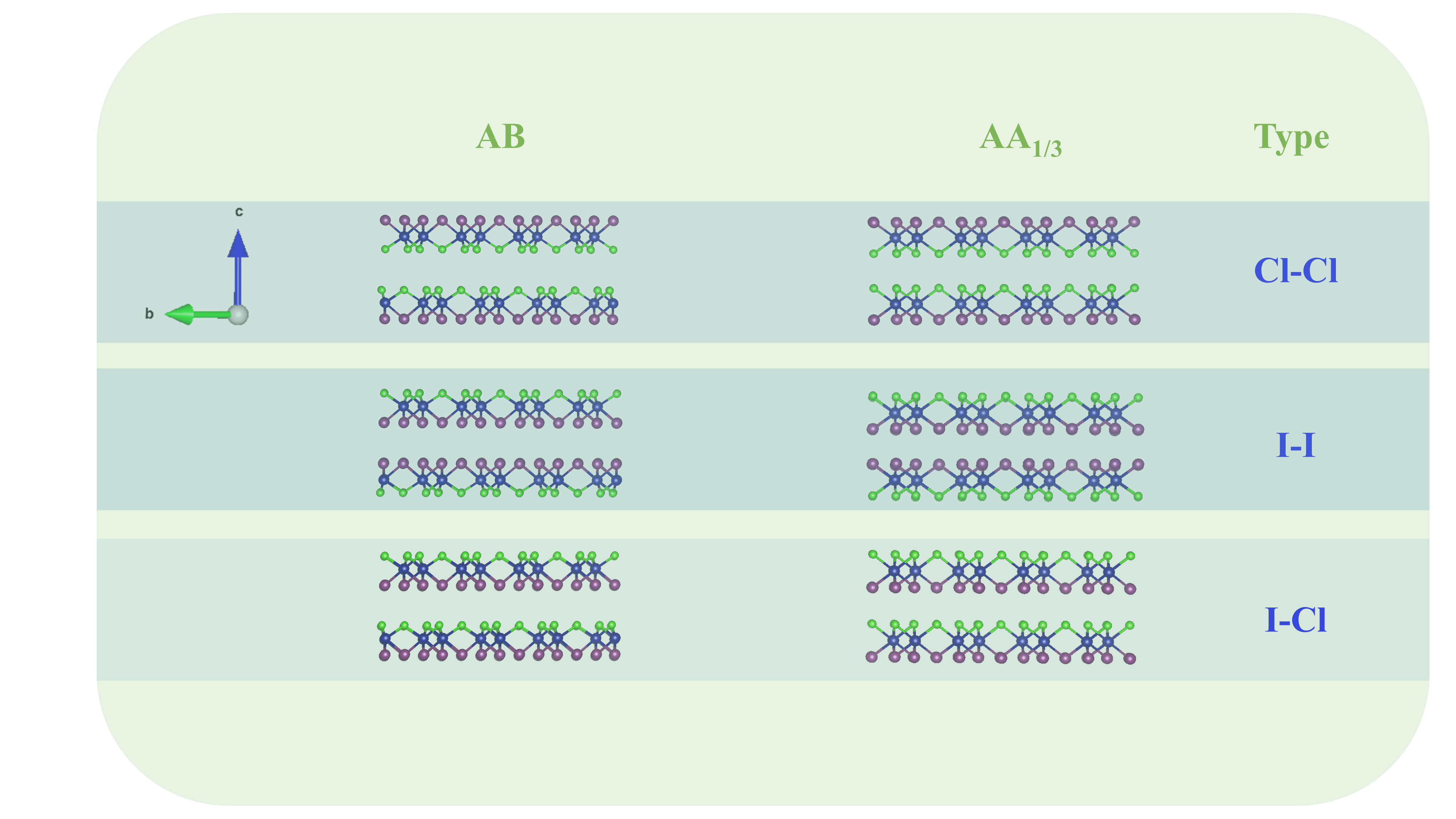}
	\caption{(Color online)
Structural models of the three configurations with AB and AA$ _{1/3} $. Purple, green, and blue respectively represent I, Cl and Cr atoms.
    \label{Figx0} }
\end{figure}

The three configurations for the Janus bilayer of Cl$_3$Cr$_2$I$_3$ with AB and AA$_{1/3}$ stacking orders are presented in Fig.~\hyperref[Figx0]{\ref*{Figx0}}, and are labeled by Cl-Cl, I-I and I-Cl, respectively. Obviously, the important difference between the AB and AA$_{1/3}$ orders is the different slippages between two layers. The AA$_{1/3}$ stacking order is that one layer slips $a$/3 along the in-plane lattice vectors ($\vec{a}$  or $\vec{b}$ ), and the AB stacking order is obtained by shifting one layer with [$\frac{2}{3}$$\vec{a}$, $\frac{1}{3}$$\vec{b}$] relative to another layer. In fact, the AB and AA$_{1/3}$ stacking orders are analogous to the low temperature and high temperature phases of bilayer CrI$_3$, respectively. Each Janus monolayer is composed of three atomic layers, forming a sandwich structure similar to MoS$_2$ monolayer.

Note that we adopt primitive cells for all the calculations rather than the supercells in Fig.~\hyperref[Figx0]{\ref*{Figx0}}. After the geometry optimizations, all the configurations with the paramagnetic (PM) states have the smallest lattice constants, while the spin polarization effects greatly expand the lattice constants (see Table \ref{tx1}). The ferromagnetic (FM) and antiferromagnetic (AFM) states have almost equal lattice constants. This implies that the interlayer effect on the lattice constants is slight weak. Thus, the differences in lattice constants caused by the stacking orders are also slight.

The SOC effects on the lattice constants are smaller than those of the spin polarization effects, which is supported by the fact that the changes of lattice constants caused by them are relatively slight. For instance, the lattice constant of the Cl-Cl configuration with the AB stacking order and FM state is about 0.04 \AA ~larger than that of the paramagnetic (PM) state, and is only 0.003 \AA ~smaller than that including the SOC effect. With the same magnetic states, the differences of lattice constants between different configurations are small and comparable to the differences caused by the SOC effect. With the same configurations, the differences of lattice constants caused by the different stacking order are also very small. Possibly, this implies that the SOC effect, the atomic configuration and the stacking order have comparable influences on the lattice constants and  thus commonly determine the ground state of the Janus bilayers.

Another important parameter related to crystal structure is the interlayer distance (ID) summarized in Table \ref{tx2}. In this work, the distance between two layers composed of Cr atoms is defined as the ID. Obviously, the spin polarization effect could expand the IDs of the I-I and Cl-I configurations, however, it shortens the ID of the Cl-Cl configuration with the AB stacking order.
For the same configurations, the IDs of the AB stacking orders are smaller than that of AA$_{1/3}$ orders. Moreover, the ID also depends on the types of adjacent atoms between layers. The IDs of the configurations I-I, Cl-I and Cl-Cl are the largest, medium and the smallest, respectively. It can reasonably be explained by the fact that the electronegativity and atomic radius of adjacent I atoms are lower and larger than these of Cl atoms. The Cl atoms with larger electronegativity and smaller atomic radius are prone to attract the metal atom Cr in the other layer, narrowing the ID. The introduction of the SOC effect can shorten the ID and thus strengthen the interlayer effect except in the I-I configuration with the $AA_{1/3}$ stacking order.
 \begin{table}[h]
\centering
\caption{\label{tx1}%
Lattice constants (in units of \AA ) of the three configurations with different stacking orders and magnetic states}
\resizebox{1.0\columnwidth}{!}{
\begin{tabular}{ccccccccc}
\\
\Xhline{1.2pt}
\multirow{2}{*}{\textbf{Magnetic state}}  &\multicolumn{2}{c}{Cl-Cl}  &\multicolumn{2}{c}{I-I} &\multicolumn{2}{c}{Cl-I} \\  

\cline{2-7}
&AB&AA$_{1/3}$&AB&AA$_{1/3}$&AB&AA$_{1/3}$\\
\Xhline{1.2pt}
Paramagnetic &	6.550	&6.551	&6.554	&6.553	&6.554	
&6.553\\
Ferromagnetic	&6.590	&6.589	&6.588	&6.587	&6.588	
&6.589\\
Ferromagnetic+SOC &6.593&	6.594	&6.586	&6.593	&6.586	&6.594\\
Antiferromagnetic	&6.589	&6.589	&6.588	&6.587	&6.588	&6.588\\
Antiferromagnetic+SOC &6.592&6.593	&6.592	&6.593	&6.592	&6.593\\


\Xhline{1.2pt}
\end{tabular}
}

\end{table}

\begin{table}[h]
\centering
\caption{\label{tx2}%
 Interlayer distances (in units of \AA ) of the three configurations with different stacking orders and magnetic states
}
\resizebox{1.0\columnwidth}{!}{
\begin{tabular}{ccccccccc}
\\
\Xhline{1.2pt}
\multirow{2}{*}{\textbf{Magnetic state}}  &\multicolumn{2}{c}{Cl-Cl}  &\multicolumn{2}{c}{I-I} &\multicolumn{2}{c}{Cl-I} \\  

\cline{2-7}
&AB&AA$_{1/3}$&AB&AA$_{1/3}$&AB&AA$_{1/3}$\\
\Xhline{1.2pt}
Paramagnetic&	6.556	&6.564	&7.794	&7.574	&6.900	&6.918\\
Ferromagnetic&	6.645	&6.655	&7.851	&7.969	&6.926	
&6.960\\
Ferromagnetic+SOC	&6.643	&6.655	&7.834	&8.011	&6.910	&6.933\\
Antiferromagnetic	&6.647	&6.658	&7.850	&8.048	&6.935	&7.236\\
Antiferromagnetic+SOC&6.645	&6.653	&7.803	&8.086	&6.910	&6.935\\


\Xhline{1.2pt}
\end{tabular}
}

\end{table}

The ID differences between the Cl-Cl and I-I configurations are more than 1.0 \AA. Interestingly, such large ID differences have negligible influence on the difference of total energy (see Table \ref{tx3}). This implies that the interlayer interaction is very weak. Without the SOC effect, the three configurations, whether AB stacking or $AA_{1/3}$ stacking orders, have the FM ground states. Considering the SOC effect, the ground states of the Cl-Cl and Cl-I configurations are AFM and FM states, respectively, both with $AA_{1/3}$ stacking. However, the I-I configuration has the ground FM states with the AB stacking. These imply that the SOC effect, the stacking order and the atomic configuration are in  competitive relationships and therefore jointly determine the magnetic ground state. Previous studies indicate the Janus Cl${_3}$Cr${_2}$I${_3}$ monolayer possesses the intrinsic dipole moment and electric polarization \cite{r15} due to the breaking of the out-of-plane mirror symmetry. Therefore, the dipole correction is applied to all the calculations in this work. It is found that the dipole correction has a significant influence on the total energy. For instance, without the dipole correction, the Cl-I configuration with the AA$_{1/3}$ and FM state has the lowest energy in the three configurations.

\begin{table}[h]
\centering
\caption{\label{tx3}%
Total energies (in units of meV, referred to the most stable configuration I-I with AB stacking order and FM states) of the three configurations with different stacking orders and magnetic states.
}
\resizebox{1.0\columnwidth}{!}{
\begin{tabular}{ccccccccc}
\\
\Xhline{1.2pt}
\multirow{2}{*}{\textbf{Magnetic state}}  &\multicolumn{2}{c}{Cl-Cl}  &\multicolumn{2}{c}{I-I} &\multicolumn{2}{c}{Cl-I} \\  

\cline{2-7}
&AB&$AA_{1/3}$&AB&$AA_{1/3}$&AB&$AA_{1/3}$\\
\Xhline{1.2pt}

Ferromagnetic&864.50&864.64&861.53&862.74&865.47
&863.77\\
Ferromagnetic+SOC&3.06&3.13&0.00&0.94&3.53
&1.97\\
Antiferromagnetic&864.64&864.68&862.63&863.55&867.42
&863.17\\
Antiferromagnetic+SOC &2.97&2.90&0.47 &1.79&6.06
&3.15\\


\Xhline{1.2pt}
\end{tabular}
}

\end{table}

Although the I-I configuration with AB stacking order has the lowest energy, it with AA$ _{1/3  } $ stacking can also be synthesized. This is because the difference between the AB and AA$ _{1/3}$ stacking orders is negligible, which is easily affected by external factors such as the preparation process and the substrate.
For example, it is verified that the bilayer CrI$ _3 $  exhibits coexistence of two stacking orders at low temperature. Furthermore, it is possible that the other configurations Cl-Cl and Cl-I will be successfully prepared with the development of two-dimensional material preparation technology, because the three configures significant structural differences and thus can be referred to as three different materials.

\subsection{Electronic and magnetic properties}

\begin{figure}[h]
\centering
\includegraphics[width=1\columnwidth]{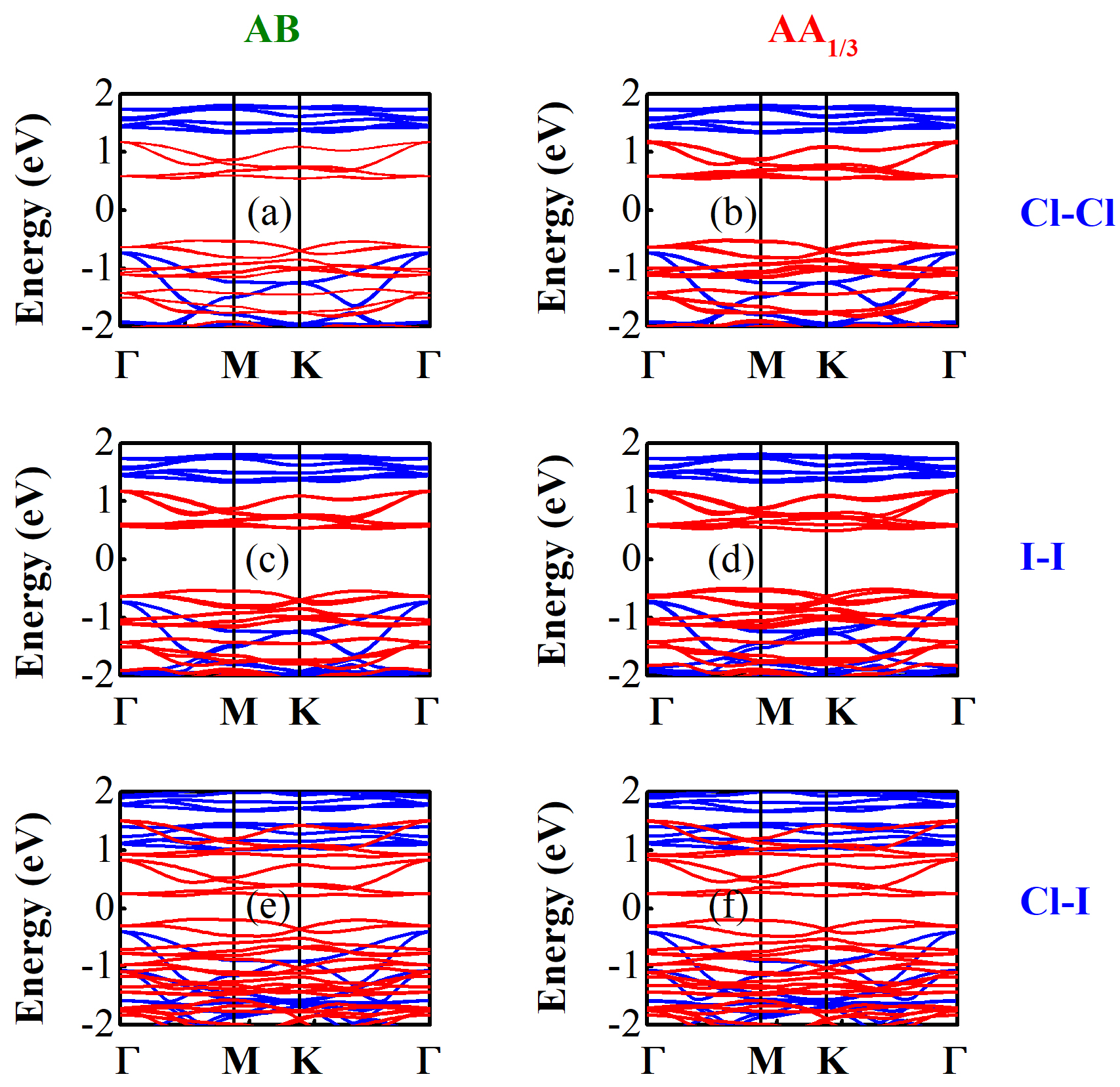}
	\caption{(Color online) The FM band structures of the three configurations (Cl-CL, I-I and Cl-I) with AB (left panel) and AA$ _{1/3 } $ (right panel) stacking orders. The red and blue represent the spin-up and spin-down states. The Fermi level is set to 0 eV.
    \label{Figx1} }
\end{figure}

\begin{figure}[h]
\includegraphics[width=1\columnwidth]{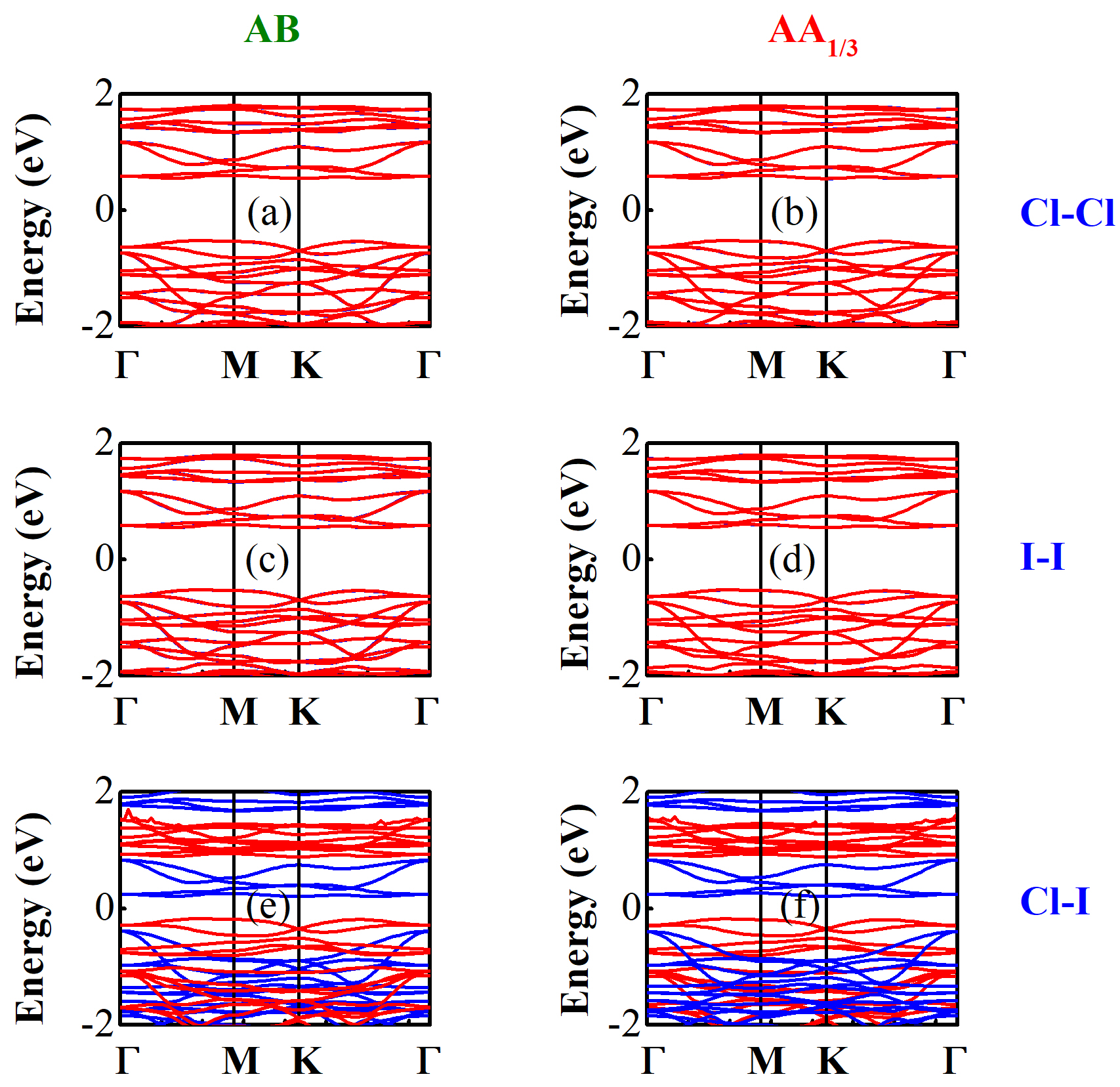}
\caption{(Color online) The AFM band structures of the three configurations (Cl-CL, I-I and Cl-I) with AB (left panel) and AA$ _{1/3 } $ (right panel) stacking orders. The red and blue represent the spin-up and spin-down states. The Fermi level is set to 0 eV.
\label{Figx2}}
\end{figure}

%

\begin{figure}[h]
\centering
\includegraphics[width=1.00\columnwidth]{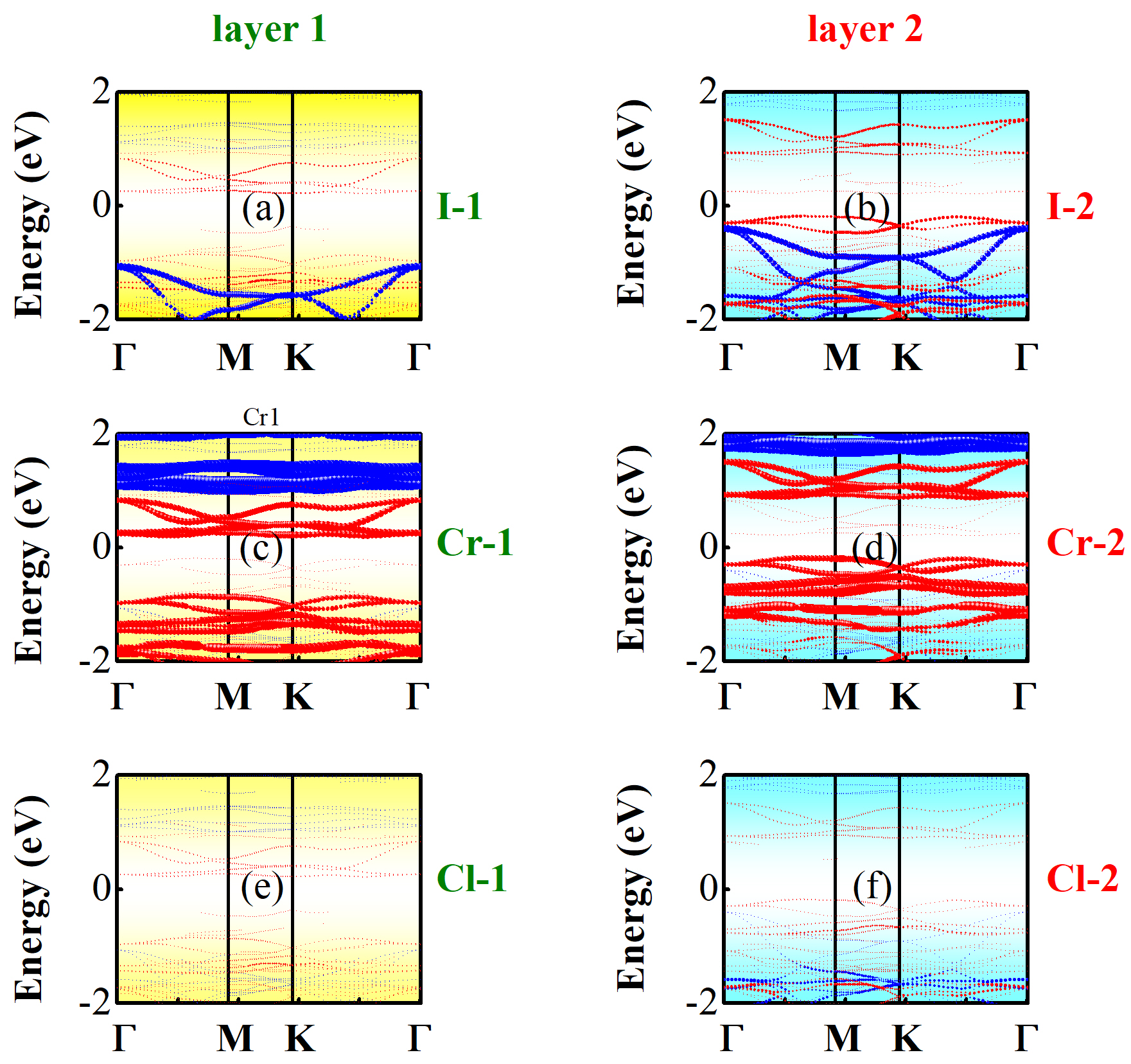}
\caption{(Color online) The projected FM band structures of the configuration (Cl-I). Band structures of layers 1 and 2
(left and right panels). I-1, Cr-1 and Cl-1 denote
the I, Cr and Cl layers of layer 1, and I-2,
Cr-2 and Cl-2 denote the I, Cr and Cl layers of layer 2, respectively. The red and blue represent the spin-up and spin-down states. The Fermi level is set to 0 eV.
\label{Figx5}}
\end{figure}

\begin{figure}[h]
\includegraphics[width=1.00\columnwidth]{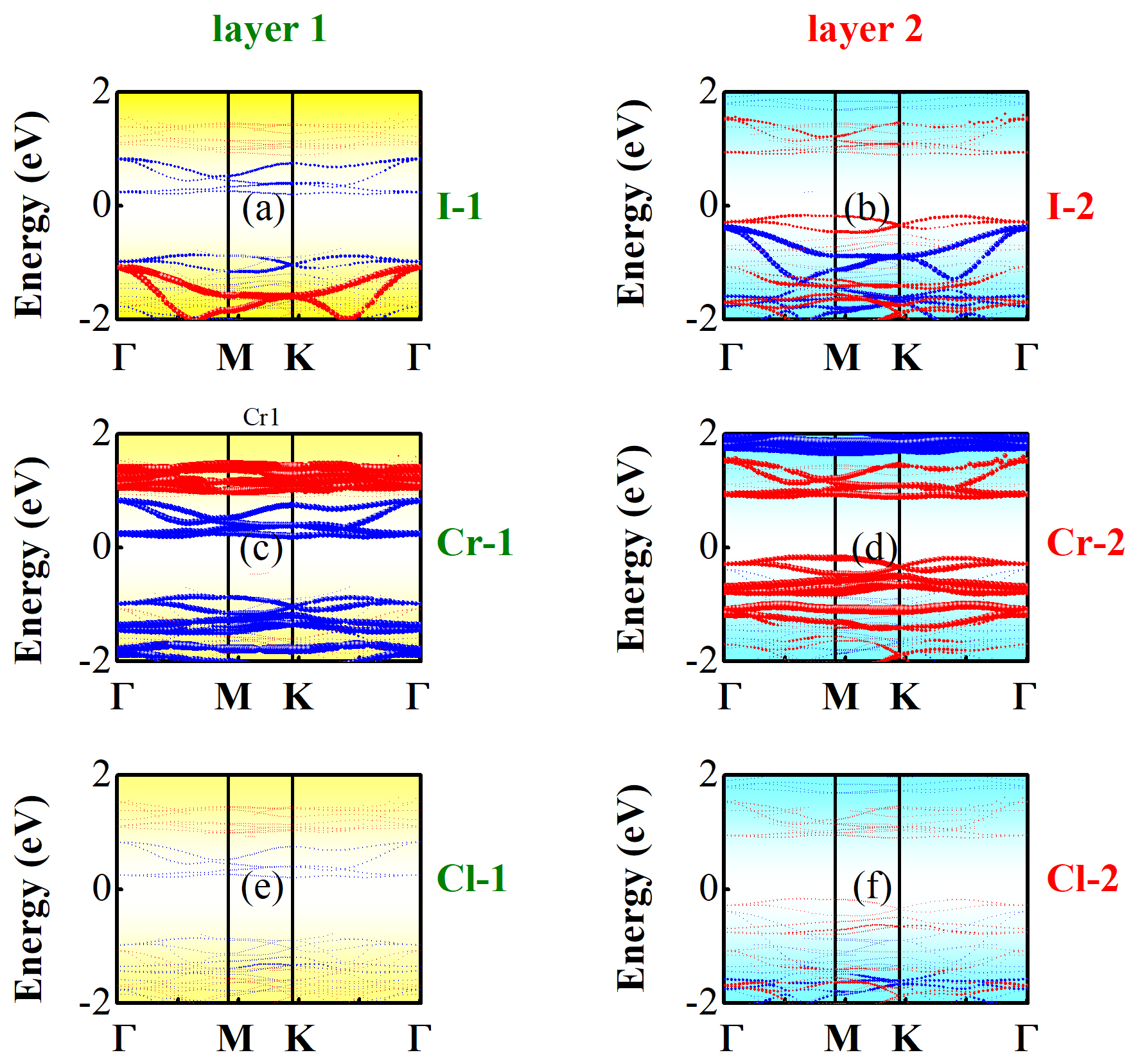}
\caption{(Color online) The projected AFM band structures of the configuration (Cl-I). Band structures of layers 1 and 2 (left and right panels). I-1, Cr-1 and Cl-1 denote the I, Cr and Cl layers of layer 1, and I-2,
Cr-2 and Cl-2 denote the I, Cr and Cl layers of layer 2, respectively. The red and blue represent the spin-up and spin-down states. The Fermi level is set to 0 eV.
\label{Figx6}}
\end{figure}

The band structures of the FM and AFM states are plotted in Fig.~\hyperref[Figx1]{\ref*{Figx1}} and Fig.~\hyperref[Figx2]{\ref*{Figx2}}, respectively. Obviously, the band gaps of the Cl-Cl and I-I configurations have comparable values and are larger than that of the Cl-I configuration. The band gap
difference between the FM and AFM states of the same configuration is negligible. This implies that the influence of the magnetic order on the band gap can be ignored. For the FM states, the band gap of the spin-up states is in the band gap of the spin-down states, thus, the value of the latter is larger than that of the former. The band structures of the Cl-Cl and I-I configurations with the $AA_{1/3}$ stacking order have obvious splitting in the spin-up states compared to that with AB stacking order (see Fig.~\hyperref[Figx1]{\ref*{Figx1}}). Especially, the splitting of the I-I configuration is as large as 0.13 eV around the high-symmetry K point.
However, the effect of stacking order on the band structure of the AFM states
is not significant or almost non-existent.

Without considering the SOC effect, the AFM state of the material usually has no spin splitting between the spin-up and spin down states. For instance, the spin-up states and the spin-down states of the Cl-Cl and I-I configurations overlap very well. Abnormally, the spin splittings of the AFM state for the Cl-I configurations are obvious. The large spin splittings are also present in the monolayer Mn$_{2}$ClF with the ground AFM state \cite{r34}. Without distinguishing between spin states, we found that the band edge shapes of FM and AFM states are similar. The difference between FM and AFM states appears in the conduction bands. It is obvious that the conduction bands of the AFM states belong to the spin-down states (see Fig.~\hyperref[Figx1]{\ref*{Figx1} (e) and (f)})and that of the FM belong to spin-up states (see Fig.~\hyperref[Figx2]{\ref*{Figx2} (e) and (f)}). This is attributed to the fact that in the AFM state, there is a layer of Cr with the opposite spin direction.

The SOC band structures of the FM and AFM are plotted in Fig.~S1 and Fig.~S2 of the Supplemental Material, respectively. Overall, the SOC effects not only cause the band splittings but also change the band shapes of the three configurations. The obvious splitting between the sixth and seventh CBs is as high as 0.3 eV at the $\Gamma$, and the VBM changes from the middle of the $\Gamma$ and K points to $\Gamma$. The SOC effect can reduce slightly the bandgaps. For instance, the Cl-I configuration with the $AA_{1/3}$ order has a SOC bandgap of 0.33 eV that is lower than 0.44 without the SOC.

Next, we focus on the  Cl-I configuration with the $AA_{1/3}$ stacking order due to its anomalous electronic structure. Its projected FM band structures are plotted in Fig.~\hyperref[Figx5]{\ref*{Figx5}}. Obviously, the highlighted band structure of the layer 1 (showed in the left panel) has relatively low energy compared to that of the layer 2 (showed in the right panel).
The energy difference between the layers 1 and 2 is about 0.7 eV. The band structures of the same elements (Cl or I) in the different layers have apparent differences, because the positions of the same atomic layers in their respective layers are not equivalent. Namely, the Cl layer in the layer 1 is far from the middle of the bilayer and that in the layer 2 is the opposite case. The I layers are also in this situation. The I layer in the layer 2 has more contribution to the VB than that in the layer 1 has. Whether in layer 1 or layer 2, the contribution of Cl layer to band edge is negligible. Additionally, the CB and VB mainly come from the contributions of the Cr layer in the layers 1 and 2, respectively. Hence, the transport properties are determined by the Cr layers. The CB of the spin-up states is mainly composed of d orbitals of the Cr atoms in the layer 1, and the VB comes from the d orbitals of the Cr atoms and the p orbitals of the I atoms in the layer 2. It is worth mentioning that the layer 1 also contributes to the VB, but it is very slight.

The AFM state of the Cl-I configuration with the $AA_{1/3}$ stacking order is unusual and interesting due to the asymmetry between the spin-up and spin-down states. Fig.~\hyperref[Figx6]{\ref*{Figx6}} indicates that the band structures of the layer 2 are similar to those of the FM states. The VB is still composed of the spin-up states of the Cr and I layers. Compared to the project FM band structure, a \textquotedbl spin flip \textquotedbl ~occurs in the  band structure in the layer 1 where the Cr atoms are set to the opposite spin. Namely, the original spin-up/down has changed to spin-down/up states. Thus, the VB mainly comes from the contribution of the spin-down states of the I layer. The AFM band structure of the Cl-I configuration with the AB order is similar to that with the AA$_{1/3}$ order. Possibly, this implies the abnormal AFM state has little or no correlation with the stacking order.

\begin{figure}[t]
\centering
\includegraphics[width=0.90\columnwidth]{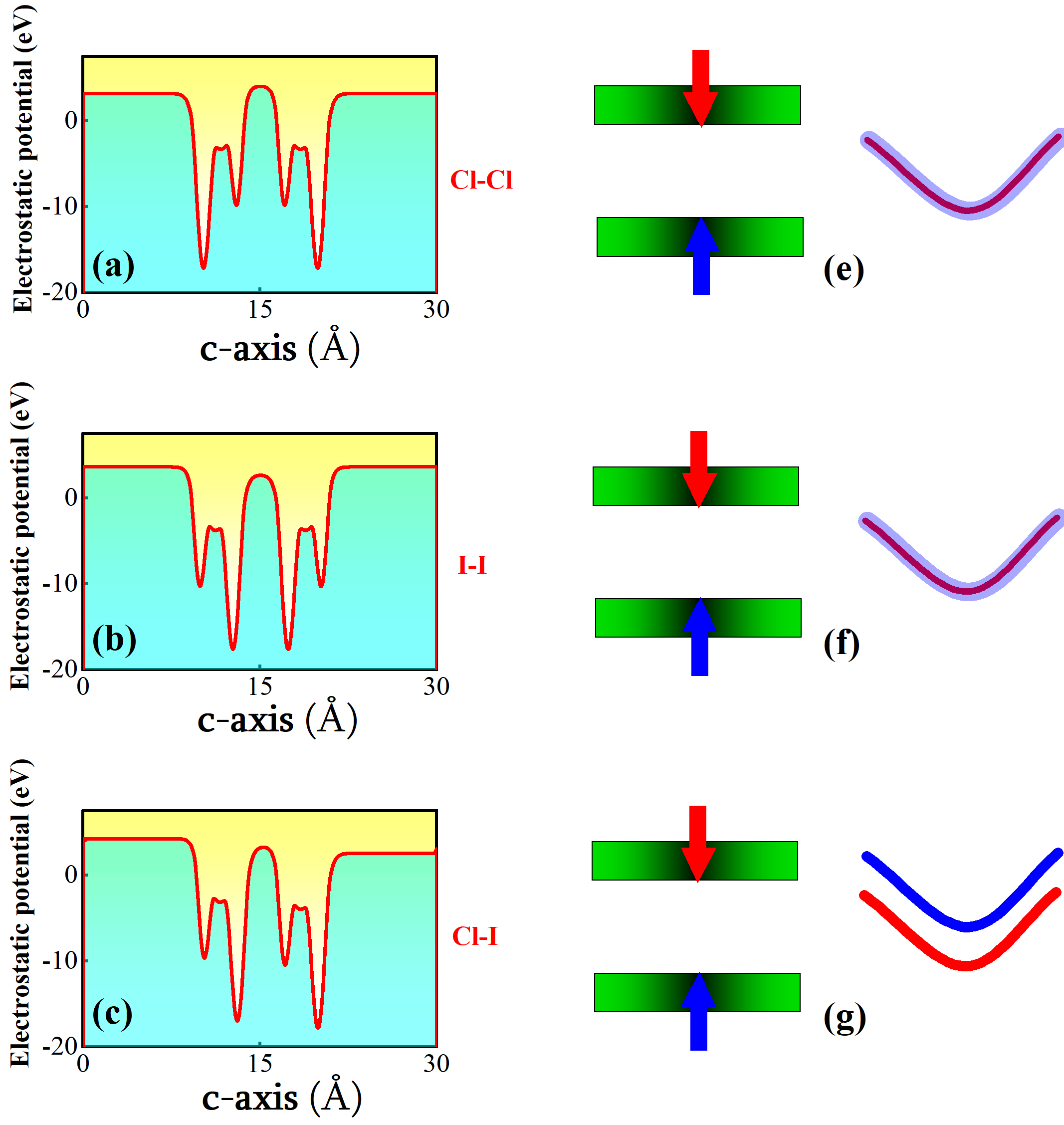}
\caption{(Color online) The average electrostatic potential of the Cl-Cl (a), I-I (b) and Cl-I (c) configurations, and corresponding antiferromagnetic configurations and schematic band diagrams (e)-(g).
\label{Figx7}}
\end{figure}

First, the abnormal AFM state is related to the symmetry breaking of the average electrostatic
potential along the c-axis direction. Fig.~\hyperref[Figx7]{\ref*{Figx7}} shows that the average electrostatic
 potentials of the I-I and Cl-Cl configurations are symmetrical about the center of the \emph{c} axis. The average electrostatic
 potential of the Cl-I configuration is higher at the Cl side than that at the I side. As a result, a vertical dipole moment pointing from I to Cl occurs. Second, the destruction of the periodic structure is also an important reason for the abnormal AFM state. To verify this, we construct a periodic structure of the Cl-I bulk, and find that the spin splitting of the AFM state is negligible. Finally, we consider that weak interlayer interactions are also an important reason for the formation of this anomalous AFM state.

As mentioned above, the bilayers do not have periodicity along the c-axis direction and have the breaking of the symmetry of the average  electrostatic potential. Hence, there are  higher electrostatic potential at the Cl side than at the I side, and the electrostatic potential difference between the two sides reaches 1.7 eV. This is inevitably accompanied by the transfer of interlayer charges. According to the Bader charge analysis \cite{r26,r27,r28,r29}, each Cl and I atoms gain 0.52 e and 0.29 e, respectively, and each the Cr atom loses 1.23 e. As a result, there is a vertical dipole moment of 3.35 D. However, the Cl-Cl and I-I configurations have most negligible vertical polarization, because the dipole moments in the two layers are exactly opposite.

Finally, we use the mean field approximation \cite{r30,r31,r32} to evaluate the critical temperatures of the Janus bilayers. The critical temperatures are defined as
\begin{equation} \label{e1}
T_c=\frac{2}{3k_B}J,
\end{equation}
where $k_B$ is the Boltzmann constant, and $J$ is the exchange parameter that can be calculated by the following formula
\begin{equation} \label{e1}
J=\frac{E_{AFM}-E_{FM}}{N|\vec{S}|^2},
\end{equation}
where $E_{AFM}$ and $E_{FM}$ are the total energies of the AFM and FM states, and $N$ is the number of the nearest neighbor atoms and $\vec{S}$ is the spin vector with $|\vec{S}|=\frac{3}{2} $. In this work, the intralayer interactions are considered and the interlayer interactions are neglected due to their weakness. Moreover, the intralayer interactions only contain the nearest neighbor interaction and the influence of the spin change on the lattice structures is ignored. For the AA$_{1/3}$ stacking order, the critical temperatures of the Cl-Cl, I-I, and Cl-I configurations are 32.6 K, 30.0 K and 31.9 K, respectively. For the AB  stacking order, they are 31.6 K, 33.5 K and 34.3 K. Obviously, the critical temperatures are lower than that of CrI$_3$ monolayer (45 K) \cite{r2} and CrI$_3$ bulk (61 K) \cite{r33} .

\section{Conclusion}
To sum up, we theoretically explore the crystal structures and electronic magnetic properties of the Janus bilayers (Cl-Cl, I-I and Cl-I configurations). The three configurations have large ID differences, however, their total energy differences are small. The magnetic ground states of the three configurations are related to the atomic configuration, the SOC effect and the stacking order. There is a vertical dipole moment and an abnormal AFM state in the Cl-I configuration, which is attributed  to non-periodic structure, symmetry breaking of the average potential and weak interlayer interaction.

\noindent{\textbf{CRediT authorship contribution statement}}\\

\noindent{\textbf{Declaration of competing interest}}\\

The authors declare that they have no known competing financial interests or personal relationships that could have appeared to influence the work reported in this paper.\\

\noindent{\textbf{Data availability}}\\

No data was used for the research described in the article.\\

\noindent{\textbf{Acknowledgement}}\\

This work is supported by the Scientific and Technological project of Nanchang Institute of Science and Technology (Grant No. NGKJ$-$22$-$09).
\\
~

\noindent{\textbf{References}}
\bibliographystyle{unsrt}
\bibliography{apsguide4-2}
\end{document}